# A unified quantum electrical platform for synchronous metrological realization of volt, ohm and ampere

Lei Wang[1,5†], Lushuai Qian[3†], Yunfeng Lu[2], Yang Shi[2], Xuanlu Yang[1], Xiaoding Huang[4], Zihan Lei[1,5], Yuan Zhang[1], Yaqiong Fu[3], Junsheng Cheng[1,5], Jianhua Liu[1,5], Xinning Hu[1,5], Yinming Dai[1,5], Jianting Zhao[2*] and Qiuliang Wang[1,5*]

[1]Institute of Electrical Engineering, Chinese Academy of Sciences, Beijing 100190, China.
[2]National Institute of Metrology, Beijing 100029, China.
[3]College of Mechanical and Electrical Engineering, China Jiliang University, Hangzhou 310018, China.
[4]Beijing Orient Institute of Measurement and Test, Beijing 100081, China
[5]University of Chinese Academy of Sciences, Beijing 100049, China

*Corresponding authors. E-mails: zhaojt@nim.ac.cn; qiuliang@mail.iee.ac.cn
†These authors contributed equally to this work.

**Abstract**

A quantum electrical standard system with co-located integration is essential for reducing reliance on distributed traceability chains in high-accuracy electrical metrology, particularly for portable and on-site applications. From the perspective of metrological completeness, the independent realization of any two of the three fundamental electrical quantities—voltage, resistance and current—is sufficient, as the third can be derived through Ohm's law. Among the available routes, the combination of Josephson voltage and quantum Hall resistance standard offers superior uncertainty performance, but is fundamentally constrained by the incompatibility between the tesla-level magnetic field required for the quantum Hall effect and the near-zero magnetic field environment necessary for stable Josephson operation. Here a compact unified quantum electrical platform is reported that enables the co-realization of quantum voltage and resistance within a single cryostat operating near 4 K, with quantum current derived within the same platform via Ohm's law. By establishing a hierarchical magnetic shielding architecture with staged attenuation and spatial field confinement, magnetic environments of 6 T and below 50 nT can coexist within an axial separation of 270 mm, with negligible cross-coupling. Under fully integrated operation, the Josephson and quantum Hall subsystems agree with their expected quantized values within relative standard uncertainties of $2.6\times10^{-9}$ and $1.4\times10^{-8}$, respectively. Meanwhile, by linking these two standards through an improved cryogenic current comparator, a quantum current of 50 μA is realized with a relative standard uncertainty of $6.6\times10^{-8}$. These results demonstrate that three basic electrical units can be synchronously realized with superior metrological consistency on a single integrated platform, offering a viable transition from distributed calibration chains toward compact-integrated quantum-based realization.

## Introduction

The implementation of the revised International System of Units (SI) in 2019 heralded a fundamental paradigm shift in modern metrology [1]. By defining all seven base units against invariant fundamental physical constants, the revised SI now provides unprecedented stability, universality and reproducibility [2]. Following the revised SI, voltage, resistance and current can be physically realized via the Josephson effect (JE) [3], the quantum Hall effect (QHE) [4] and single-electron transport (SET) [5], respectively, collectively forming the cornerstone of modern electrical metrology [6-8]. From a metrological system perspective, a complete quantum electrical metrology framework does not require independent realization of all three quantities: implementation of any two is sufficient, with the third deterministically derivable through Ohm's law. For instance, quantum current can be derived from quantum voltage and resistance [9-12]. This intrinsic flexibility enables multiple viable technical routes for constructing a system-level quantum electrical metrology platform.

Quantum electrical standards have undergone decades of evolution toward compact, application-driven metrological systems since the 1970s. Programmable Josephson voltage standards (PJVS) [13-19], together with Josephson arbitrary waveform synthesizers (JAWS) [20-28], deliver spectrally pure DC and AC waveform synthesis at sub-nanovolt precision within cryogen-free systems [29-32], laying a robust technical foundation for quantum-traceable voltage [33-46], power and energy [47-54], as well as ratio and impedance standards [55-62]. In terms of resistance metrology, GaAs/AlGaAs-based quantum Hall resistance (QHR) devices have long served as primary standards with relative uncertainties down to the $10^{-11}$ level [63-68]. Their widespread practical uptake, however, is severely hampered by extreme operating requirements: a magnetic field of approximately 10 T and sub-kelvin cryogenic temperatures. Graphene-based QHR devices have emerged as a transformative alternative, retaining metrological accuracy at the $10^{-9}$ level under more relaxed operating conditions (3–5 T magnetic field, temperature ≥4 K) while offering enhanced current-carrying capacity exceeding 200 μA [69-86]. These merits greatly facilitate the development of portable and cryogen-free QHR standards. By comparison, the maturation of quantum current realization has lagged substantially behind voltage and resistance quantum standards. Single-electron transport architectures are restricted to picoampere-scale output currents with relative uncertainties ranging from $10^{-6}$ to $10^{-7}$ [87-96]. In contrast, Ohm's-law-traceable schemes combining PJVS and QHR, aided by cryogenic current comparators (CCC), can generate currents spanning the nanoampere to milliampere range with uncertainties better than $10^{-8}$ [9-12, 97].

In current metrological practice, individual quantum electrical standards are housed in discrete systems and interconnected via hierarchical calibration traceability chains. While this distributed paradigm delivers exceptional measurement accuracy and long-term stability and remains the technical backbone of modern electrical metrology, it imposes inherent limitations on system integration, operational efficiency and on-site deployability. Reliance on multi-stage calibration workflows introduces additional uncertainty propagation pathways, restricting the deployment of quantum metrology outside well-controlled laboratory environments. In this context, the development of co-located, unified quantum electrical platforms—in which multiple primary quantum standards operate within a single cryogenic system—has emerged as an essential direction for advancing electrical metrology toward miniaturized, efficient, and application-oriented implementations [98].

Achieving such high-level integration requires overcoming a set of mutually coupled technical constraints, dominated by mutually exclusive environmental requirements among distinct quantum standards. Specifically, quantization of the QHR standard relies on multi-tesla magnetic fields to establish discrete Landau levels, whereas Josephson voltage standards (JVS) require nanotesla-scale residual magnetic environments to suppress flux trapping and sustain robust quantum phase coherence. This disparity spans more than eight orders of magnitude in magnetic field strength. Traditional schemes that rely exclusively on physical spatial isolation are unable to fulfill the strict requirements for compact footprint, cryogenic-system compatibility and portability, forming a fundamental barrier to the unified operation of QHR and JVS subsystems.

To circumvent this incompatibility, previous efforts have explored diverse integration strategies. Hohls *et al*. demonstrated an on-chip integration of QHR and SET devices, enabling quantized voltage generation without Josephson junctions [99]. This pioneering study established the possibility of constructing a complete quantum electrical standard system on a single chip, representing a forward-looking paradigm for miniaturized quantum metrology hardware. Nevertheless, the scheme is currently restricted to microvolt-scale outputs with relatively elevated uncertainties, inherent to the device architecture and finite single-electron transfer efficiency. More recently, Rodenbach *et al*. proposed an elegant co-integration architecture combining a quantum anomalous Hall resistor (QAHR) and PJVS in a single cryostat [100]. Benefiting from the zero-field quantization property of QAHR, the system eliminates high magnetic field requirements and enables quantum current generation ranging from 9.33 to 252 nA with relative uncertainties down to the $10^{-6}$ level, where the primary uncertainty arises from microwave excitation of PJVS interfering with QAHR quantization. Previous studies have confirmed that QAHR holds the potential to match the accuracy of conventional QHR standards [101-103], rendering the QAHR-PJVS integration a promising miniaturization solution for quantum electrical metrology. Despite these encouraging progresses, substantial gaps still exist in developing a practical platform capable of co-locating multiple quantum standards with metrological accuracy under accessible operating conditions. State-of-the-art SET and QAHR prototypes generally rely on millikelvin-scale dilution refrigeration, which imposes substantial overhead in system complexity, maintenance cost, and operational robustness, severely limiting on-site applicability.

Here, a unified quantum electrical platform is presented that co-integrates a conventional PJVS and a graphene-based QHR standard within a single cryostat operated near 4 K. Through synergistic co-design of magnetic shielding and thermal management, simultaneous quantum voltage and resistance realization is achieved within a fully unified operating environment. A hierarchically structured magnetic shielding system with graded field attenuation and spatial magnetic confinement is engineered to accommodate tesla-level and nanotesla-level magnetic domains in a compact footprint, while fully preserving the quantized behavior and metrological performance of both subsystems. Quantum current is further derived on the same platform via an optimized CCC. This work establishes a practical and scalable pathway for the co-located realization of mature Josephson and quantum Hall electrical standards, laying a critical foundation for next-generation compact, transportable quantum metrology systems.

## System design and integration strategy

### System-level constraints and coupling framework

A unified quantum electrical standard relies on the co-integration of quantum voltage and quantum resistance subsystems within a shared cryogenic environment. However, their distinct operating principles impose mutually exclusive physical boundary conditions, giving rise to multi-physics coupling and systematic performance trade-offs. Resolving such inherent incompatibility is essential for developing compact, transportable quantum metrology systems that transcend conventional standalone laboratory instruments.

These conflicting physical interactions originate from three dominant cross-domain coupling pathways: magnetic, thermal, and electromagnetic. Collectively, these pathways restrict the achievable background magnetic field level, thermal stability, and electromagnetic isolation of the integrated platform. As illustrated in Fig. 1, such multi-physics coupling fundamentally complicates the direct co-location of PJVS and QHR modules within a confined cryostat volume.

To address this challenge, the system design is formulated as a constraint-driven decoupling problem. Key coupling pathways are selectively suppressed via customized structural optimization and physical isolation schemes, allowing each quantum subsystem to operate within its inherent physical boundaries while retaining the compact form factor of the overall platform. Detailed implementation procedures and experimental validation results are provided in the following subsections.

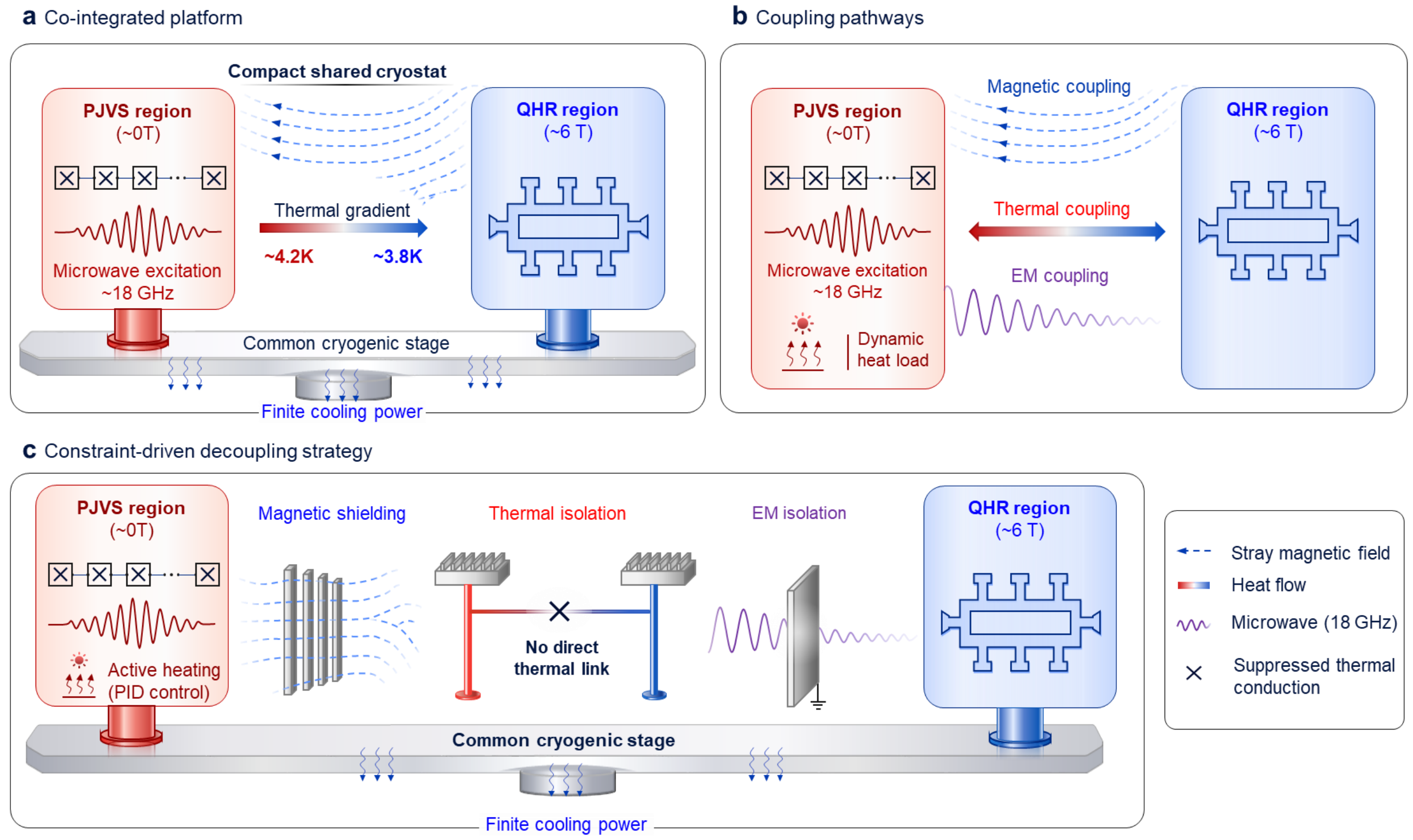

**Fig. 1 | System-level coupling pathways and constraint-driven decoupling framework for a unified quantum electrical platform**. **(a)** Co-integrated quantum electrical platform comprising a PJVS and a QHR within a shared cryogenic environment. The PJVS operates under low magnetic field with microwave excitation, whereas the QHR requires a strong static magnetic field. **(b)** Cross-domain coupling pathways between subsystems, including magnetic coupling from stray fields, thermal coupling through the common cryogenic stage with finite cooling power, and electromagnetic coupling associated with microwave excitation and leakage. **(c)** Constraint-driven decoupling strategy. Magnetic shielding, thermal anchoring and electromagnetic isolation are implemented to suppress dominant coupling pathways, enabling concurrent operation within a compact integrated platform.

## Magnetic field management through multi-stage shielding

The most stringent system constraint arises from the tesla-scale static magnetic field required for QHR Landau level quantization, which must be attenuated to the nanotesla level to protect the PJVS from magnetic flux trapping and quantum phase decoherence. This requires extreme stray-field suppression within a limited spatial footprint.

Magnetic isolation is implemented via a multilayer cascaded shielding architecture that integrates flux guiding, magnetic absorption, and flux expulsion mechanisms. Stray fields generated by the superconducting magnet are first diverted and partially attenuated by outer ferromagnetic shells with low-reluctance return paths. Intermediate high-permeability alloy laminates further suppress residual stray flux, while an innermost closed superconducting enclosure expels remnant magnetic fields through the Meissner effect.

This multi-mechanism cascaded attenuation avoids the saturation and performance limitations inherent to single-layer shielding schemes. In practice, the magnetic field is reduced from 6 T at the QHR device to below 50 nT in the PJVS working region over a horizontal distance of 270 mm, achieving a total attenuation exceeding eight orders of magnitude (Extended Data Fig. 1). Both finite-element simulations and experimental measurements confirm that the ultra-low-background magnetic environment remains stable during full-scale operation of the QHR superconducting magnet.

Beyond magnetostatic shielding, the closed enclosure also suppresses high-frequency electromagnetic crosstalk originating from PJVS microwave excitation. Although no direct experimental evidence currently confirms microwave leakage-induced perturbation of graphene-based QHR quantization, mitigating this potential interference pathway is critical for ensuring long-term system stability and metrological robustness [100]. Full-wave electromagnetic simulations show that the microwave transmission coefficient between the PJVS and QHR regions remains below −180 dB across the 16–20 GHz band (Extended Data Fig. 3), demonstrating negligible electromagnetic leakage within the shared cryogenic structure.

**Thermal decoupling under co-integrated cryogenic operation**

Thermal crosstalk represents the second major obstacle to stable synchronous operation of the co-integrated subsystems. While both PJVS and QHR devices operate near 4 K, they introduce distinct local heat loads arising from microwave dissipation, bias-current Joule heating, and structural thermal conduction. Uncontrolled thermal diffusion across the shared cold stage degrades temperature stability and impairs quantized performance.

To minimize inter-module thermal coupling, a conduction-cooled cryostat with segmented thermal conduction networks is adopted. Independent thermal anchoring and intermediate heat sinks localize heat dissipation and reduce parasitic thermal transfer between subsystems. For the PJVS subsystem, closed-loop PID temperature regulation is implemented to compensate for dynamic thermal disturbances.

Under steady-state integrated operation, both subsystems maintain stable temperatures below 4 K. Compared with standalone operation, the modified thermal boundary reduces the usable PJVS bias current margin from ~2.4 mA to ~1.4 mA (see Methods). This reduction originates from global thermal regulation constraints rather than direct heat load transfer from the QHR subsystem. Despite this moderate operational trade-off, the remaining bias margin is sufficient to support stable, repeatable quantized voltage output for metrological application.

**System-level co-integration and validation**

Based on the integrated magnetic, thermal, and electromagnetic decoupling infrastructure, a 2 V PJVS chip and a graphene-based QHR device are fully co-located within a single conduction-cooled cryogenic platform (see Methods).

Under fully synchronized operating conditions, including simultaneous strong magnetic field biasing and high-frequency microwave excitation, both subsystems exhibit stable, reproducible quantization behavior. The PJVS produces well-resolved voltage steps with reliable operational margins, while the graphene QHR maintains flat resistance plateaus with negligible longitudinal dissipation.

The absence of measurable performance degradation during concurrent operation verifies effective suppression of multi-physics cross-coupling. This validates that the proposed system-level design enables reliable co-integration of Josephson and quantum Hall electrical standards while preserving the strict physical boundary conditions required for high-precision metrological operation. Overall, this constraint-adaptive decoupling strategy provides a scalable technical route toward compact, next-generation transportable quantum electrical standard systems.

## Experimental validation

### Magnetic and thermal compatibility

Following system cooldown, both the Josephson and quantum Hall subsystems reach steady thermal stability below 4 K. During subsequent operation, the Josephson subsystem is actively stabilized at a specific set point ranging from 3.8 K to 4.2 K, whose exact value is optimized according to the real-time operating condition of the PJVS chip, with temperature fluctuations strictly confined within ±50 mK. When the superconducting magnet is fully energized to 6 T, SQUID-based measurements resolve a magnetic field variation of approximately 5 nT within the PJVS functional region. This negligible magnetic shift confirms that the ultra-low magnetic background essential for reliable Josephson quantization is well preserved under full-strength magnetic excitation, consistent with the shielding performance characterized in Extended Data Fig. 1.

### Quantum voltage accuracy under integration

The voltage quantization accuracy of the integrated platform is evaluated via differential comparison with an independent reference quantum voltage system (see Methods). At a nominal output of 2 V, ten repeated measurement trials (Extended Data Fig. 4) yield a mean voltage deviation of -0.7 nV. The corresponding combined relative standard uncertainty for this operating condition is $2.6 \times 10^{-9}$ ($k = 1$), with detailed uncertainty components summarized in Table 1. These results demonstrate that the integrated platform provides stable phase-locking conditions for reliable quantized voltage generation.

**Table 1** Uncertainty components for quantum voltage generation

| Uncertainty component | Type | Contribution($\times 10^{-9}$) |
|---|---|---|
| Measure dispersion of $V_{\mathrm{diff}}$ | A | 1.3 |
| DVM gain and linearity error | B | <1 |
| Uncertainty of the referenced PJVS system | B | <2 |
| **Combined** | | 2.6 ($k$=1) |

### Quantum resistance accuracy under integration

Quantum Hall performance is characterized under fully integrated operating conditions, with continuous microwave excitation applied to the Josephson subsystem. At a magnetic field near 6 T and a stabilized temperature of 3.8 K, a well-defined $\nu = 2$ quantization plateau is clearly resolved. The longitudinal resistance is maintained below 1 mΩ, verifying near-dissipationless quantum transport of the graphene quantum Hall device.

To further verify metrological reliability, an indirect comparison is conducted between the integrated quantum Hall device and the national primary quantum Hall resistance standard. The comparison is performed via a CCC with a calibrated 100 Ω transfer resistor (see Methods). The resistance deviation obtained from the two independent traceability pathways is $8.0 \times 10^{-9}$, with a combined relative standard uncertainty of $1.4 \times 10^{-8}$ ($k = 1$), as summarized in Table 2. This favorable consistency demonstrates that the intrinsic quantization accuracy of the quantum Hall device is well retained under integrated co-operation conditions.

**Table 2** Uncertainty components for quantum Hall resistance generation

| Uncertainty component | Type | Contribution (×$10^{-9}$) |
|---|---|---|
| Measure dispersion of resistance ratio | A | 13.2 |
| Ratio error of CCC | B | 0.2 |
| Uncertainty of the 100 Ω transfer resistor | B | 5.0 |
| **Combined** | | 14.2 ($k$=1) |

### Quantum current derived via Ohm's law

Consistent and accurate quantization of voltage and resistance allows deterministic quantum-current generation from Ohm's law. In the practical setup, the PJVS and QHR devices are interconnected through a standardized triple-connection configuration, paired with an optimized CCC that suppresses parasitic series resistance effects to a negligible level [97]. The resultant quantum current is converted to a measurable voltage signal across a calibrated 10 kΩ resistor and acquired by a high-precision digital voltmeter.

Using this configuration, a quantum current of 50 μA is realized with a relative standard uncertainty of 6.6 × $10^{-8}$ ($k$ = 1). The self-consistent measurement results of quantum voltage, resistance, and current verify the capability of the proposed platform to realize three core electrical quantum units within a single compact system, eliminating the need for intermediate multi-stage calibration transfer.

**Table 3** Uncertainty components for quantum current generation

| Uncertainty component | Type | Contribution (×$10^{-8}$) |
|---|---|---|
| Measure dispersion of synthetic current | A | 5.3 |
| DVM gain and linearity error | B | 3.8 |
| Uncertainty of the 10 kΩ transfer resistor | B | 0.5 |
| Ratio error of CCC | B | 0.02 |
| **Combined** | | 6.6 ($k$=1) |

## Conclusion

This work demonstrates a unified quantum electrical platform capable of concurrent realization of the volt, ohm, and ampere within a single 4 K cryostat. The core system incompatibility originating from coexisting tesla-scale and nanotesla-scale magnetic boundary conditions is resolved via a hierarchical multi-stage shielding design, enabling physical compatibility for the co-located operation of discrete quantum electrical standards. Complemented by systematic thermal and electromagnetic co-optimization, the proposed architecture supports synchronous operation of Josephson and quantum Hall devices with well-preserved quantization performance. The integrated system achieves quantum voltage and resistance uncertainties at the $10^{-9}$ and $10^{-8}$ levels, respectively, and delivers deterministic quantum current generation with a relative standard uncertainty of 6.6 × $10^{-8}$. These results validate the synchronous realization of three core electrical quantum units on a single compact platform. This study provides a feasible technical pathway for on-site deployable quantum metrology systems, facilitating the transition from conventional distributed calibration hierarchies toward unified, integrated quantum electrical standardization.

## Methods

### System co-design and cryogenic architecture

The structural layout and physical implementation of the unified quantum electrical platform are illustrated in Fig. 2. The system is built upon a conduction-cooled cryostat driven by a two-stage Gifford–McMahon cryocooler. A copper thermal shield is installed between the low-temperature working zone and the vacuum vessel to block radiative heat influx from room temperature. The first and second cryocooler stages are thermally anchored to the thermal shield and low-temperature core components, respectively, providing cooling capacities of 35 W at 50 K and 1.5 W at 4.2 K. The PJVS and QHR operating regions are mechanically supported by glass–epoxy struts with low thermal conductivity to minimize parasitic heat loading. To enable co-located operation of Josephson and quantum Hall subsystems within a compact footprint, a vertically partitioned configuration is adopted with a horizontal spatial separation of 270 mm. The overall system occupies a compact volume with 640 mm diameter and 750 mm height. Independent high-purity copper thermal conduction pathways and intermediate heat sinks are implemented to suppress inter-module thermal crosstalk. The total heat load at the low-temperature stage is stabilized below 0.95 W to guarantee continuous and reliable system operation.

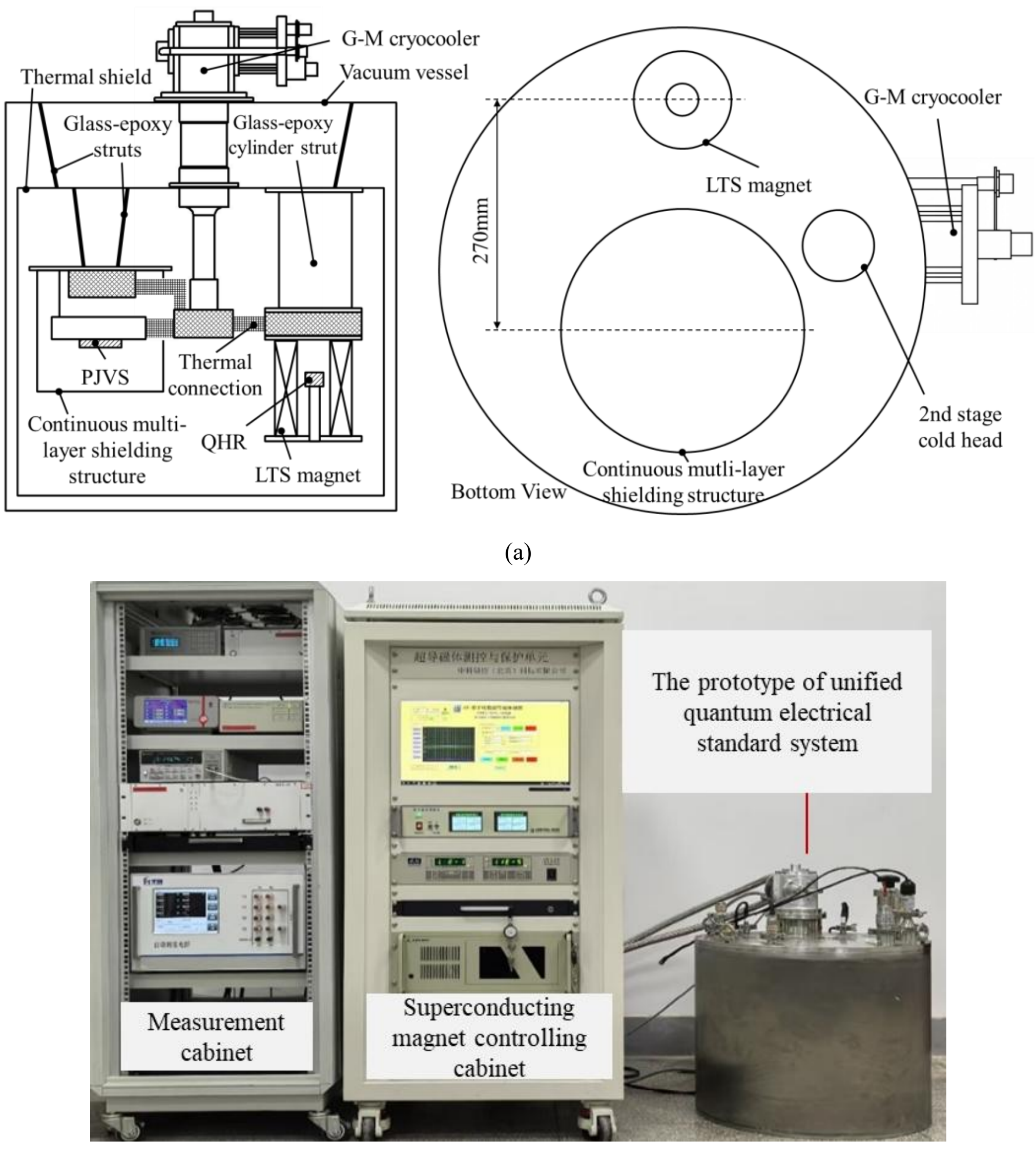

**Fig. M1 | System co-design architecture and physical photograph of the unified quantum electrical standard platform. (a)** Schematic plan layout of the integrated cryogenic assembly, illustrating the spatial arrangement of PJVS and QHR subsystems within the 4 K conduction-cooled cryostat. Key functional components, including the superconducting magnet module, magnetic shielding structure, electrical interconnections, and thermal anchoring stages, are labeled. **(b)** Photograph of the assembled platform, demonstrating the compact monolithic integration of the co-designed cryogenic and magnetic shielding architecture.

## Superconducting magnet and multilayer magnetic shielding

A multi-objective optimized magnetic shielding strategy is implemented to resolve the intrinsic conflict between tesla-scale high-field operation for QHR quantization and nanotesla-scale low-field ambient requirements for PJVS operation within a single compact cryostat [105-115]. This cascaded shielding scheme enables compact system integration and reduces sensitivity to operational sequences and external magnetic disturbances, thereby improving overall environmental robustness.

To satisfy the high-field requirements of the graphene QHR device, a customized conduction-cooled low-temperature superconducting (LTS) magnet is designed to generate stable, uniform static magnetic fields. The magnet is optimized to produce a central field of no less than 6 T within a cold bore diameter of 18 mm, providing sufficient mechanical clearance for device installation. The magnetic field uniformity is maintained within ±0.05% across a 10 mm diameter spherical volume (DSV) where the QHR chip is positioned (Extended Data Fig. 2). A hybrid linear–nonlinear optimization algorithm is adopted to determine the optimal coil geometry and shielding configuration [105-107, 111].

The LTS magnet consists of two coaxially wound NbTi superconducting coils assembled on a cylindrical mandrel and connected in series via superconducting joints. The magnet operates at a nominal current of 49.141 A with a total inductance of 3.39 H, corresponding to a stored magnetic energy of 4.1 kJ. After full energization, the magnet is switched to persistent current mode via a superconducting switch to ensure long-term magnetic field stability. Detailed geometrical and electrical parameters of the superconducting magnet are summarized in Table 4.

Hierarchical magnetic compatibility is realized via a concentric multilayer shielding stack comprising outer ferromagnetic layers, an intermediate high-permeability permalloy layer, and an inner superconducting niobium layer. The outer ferromagnetic shells divert and attenuate stray magnetic flux originating from the superconducting magnet [107, 110, 111]. The intermediate permalloy layer further suppresses residual stray fields through flux concentration and shunting [116]. Operating below its critical temperature, the innermost niobium superconductor expels remnant magnetic flux via the Meissner effect [117-120]. This multi-mechanism cascaded shielding architecture achieves progressive, eight order of magnitude magnetic field attenuation, enabling coexistence of tesla-level and ultra-low magnetic field regions within a confined space. Finite-element simulations and cryogenic experimental measurements validate the field distribution and shielding performance (Extended Data Fig. 1).

**Table 4** Parameters of low temperature superconducting magnet

| Parameters | Value |
|---|---|
| Inner diameter of superducting coil /mm | 25.2 |
| Outer diameter of superconducting coil/mm | 91.76 |
| Height of superconducting coil/mm | 160 |
| Operating current (A) | 49.141 |

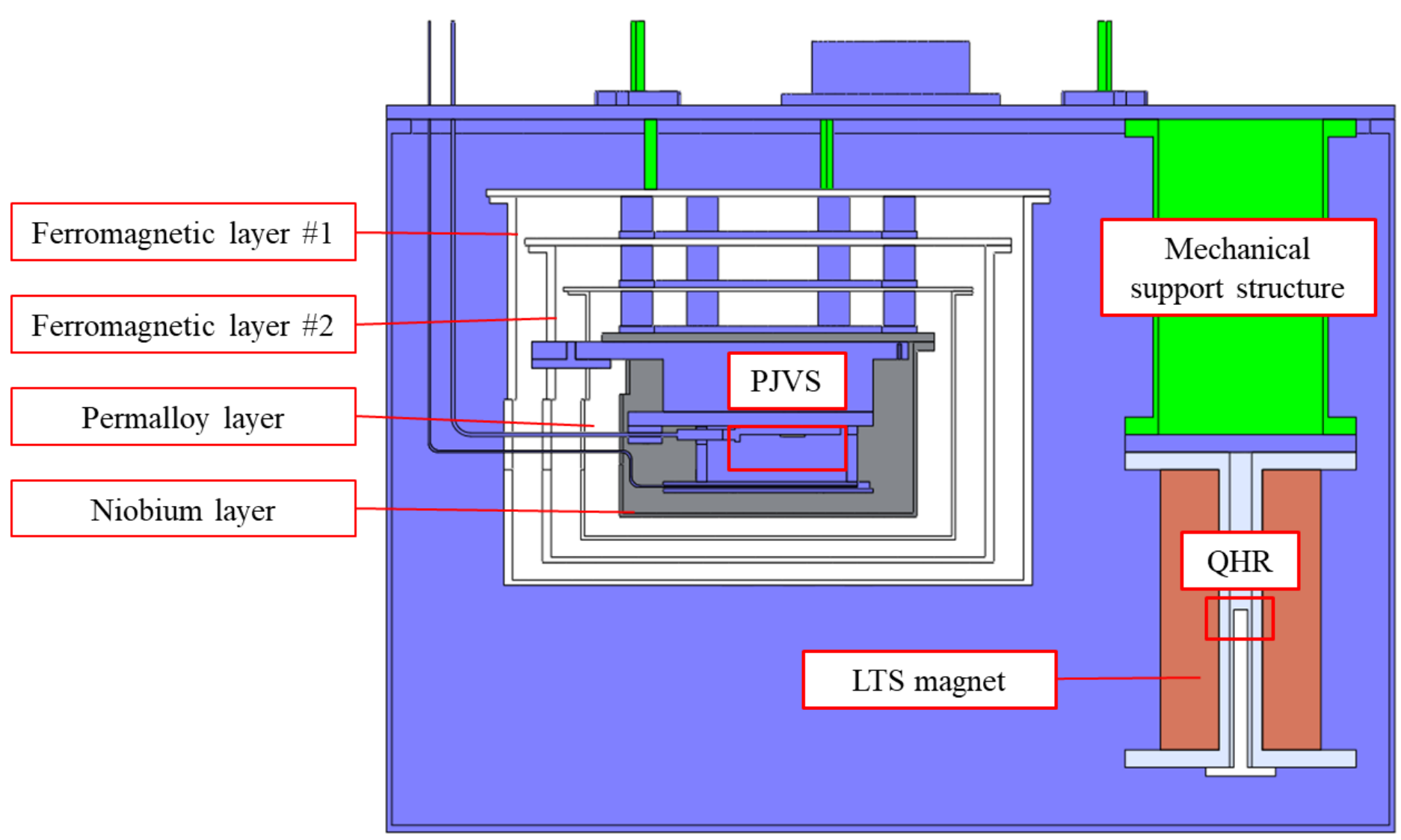

**Fig. M2 | Cross-sectional geometry of the multilayer magnetic shielding and superconducting magnet assembly.** Schematic cross-section of the integrated shielding structure and magnet configuration. The concentric shielding stack comprises two outer ferromagnetic layers, one intermediate permalloy layer, and one inner superconducting niobium layer, with respective thicknesses of 5 mm, 5 mm, 2 mm, and 5 mm. Sidewall openings are reserved for microwave transmission and electrical wiring access. The QHR device is positioned at the central homogeneous field region of the NbTi magnet, while the PJVS subsystem is enclosed within the ultra-low-field shielding cavity. The horizontal separation between the two functional regions is 270 mm.

### Quantum devices and operation methodology

The integrated quantum voltage subsystem employs a programmable Josephson junction array (PJJA) chip containing over 60,000 $Nb/Nb_xSi_{1-x}/Nb$ junctions arranged in a stacked-series configuration and partitioned into 14 independently addressable subarrays. Under 18.15 GHz microwave excitation, the chip generates a maximum quantized output voltage of ~2.3 V with a minimum programmable resolution of 75 μV. The device is driven by a dedicated control system consisting of multi-channel bias current sources, phase-locked microwave synthesizers, and automated operational software, consistent with previous established configurations [50]. The PJJA chip operates stably at a nominal temperature of 4.2 K. Quantized voltage stability is characterized via flat-spot testing [121], in which alternating subarrays are biased at the centers of the +1 and –1 quantized steps, and controlled current perturbations are applied to quantify the stable operational bias margin.

The QHR device is fabricated based on epitaxial monolayer graphene grown on a SiC substrate, exhibiting a carrier mobility exceeding 10,000 $cm^2 \cdot V^{-1} \cdot s^{-1}$. The graphene film is patterned into a standard Hall-bar geometry (100 μm × 200 μm), and Ti/Au electrodes are deposited to form low-resistance ohmic contacts. During operation, the device is subjected to a perpendicular static magnetic field generated by the LTS magnet, with the stage temperature stabilized below 4 K. Electrical characterization is performed using a precision current source and a high-sensitivity nanovoltmeter to acquire longitudinal resistance ($R_{xx}$) and Hall resistance ($R_{xy}$) , enabling quantitative evaluation of quantization quality and residual dissipation.

### Accuracy evaluation and uncertainty analysis methodology

The quantized voltage accuracy of the integrated PJVS is evaluated through differential comparison with a reference SRI-6000 PJVS system. The negative terminals of the two systems are electrically shorted, while the positive terminals are connected to an keysight 34420A nanovoltmeter for differential voltage acquisition. Measurements are performed with fixed power line cycle (NPLC) settings and repeated sampling to suppress random noise. To eliminate thermal electromotive forces (EMF) and instrumental offsets, measurements are conducted under both positive and negative output polarities, and the final differential voltage is determined via polarity-reversal averaging. Uncertainty components are assessed into Type A and Type B evaluations. The Type A uncertainty is derived from statistical analysis of repeated measurements. Type B contributions include nanovoltmeter gain and nonlinearity, and the combined uncertainty of the reference PJVS system.

The accuracy of the integrated QHR device is validated by comparison with the national primary quantum Hall resistance standard. The comparison is implemented via a calibrated 100 Ω transfer resistor and a CCC. Resistance ratio measurements are acquired under stabilized magnetic field and temperature conditions, with repeated sampling adopted for Type A statistical uncertainty evaluation. Type B uncertainty contributions originate from CCC proportional error and the combined uncertainty of the 100 Ω transfer resistor.

Quantum current is deterministically synthesized based on Ohm's law via CCC-based linkage of quantum voltage and resistance standards. Current generation is realized on the CCC primary winding, and the secondary output current is measured under a 1:1 turns ratio. The secondary quantum current is converted into a measurable voltage signal across a calibrated 10 kΩ resistor and sampled using an keysight 3458A high-precision digital multimeter. A polarity-reversal measurement scheme is applied to suppress parasitic EMF and systematic instrumental offsets. The Type A uncertainty is calculated as the relative standard deviation of the mean from consecutive voltage sampling. Type B uncertainty sources include CCC ratio inaccuracy, combined uncertainty of the 10 kΩ transfer resistor, and multimeter gain stability, which is quantified through repeated PJVS-based calibration trials.

## Extended Data Figure Captions

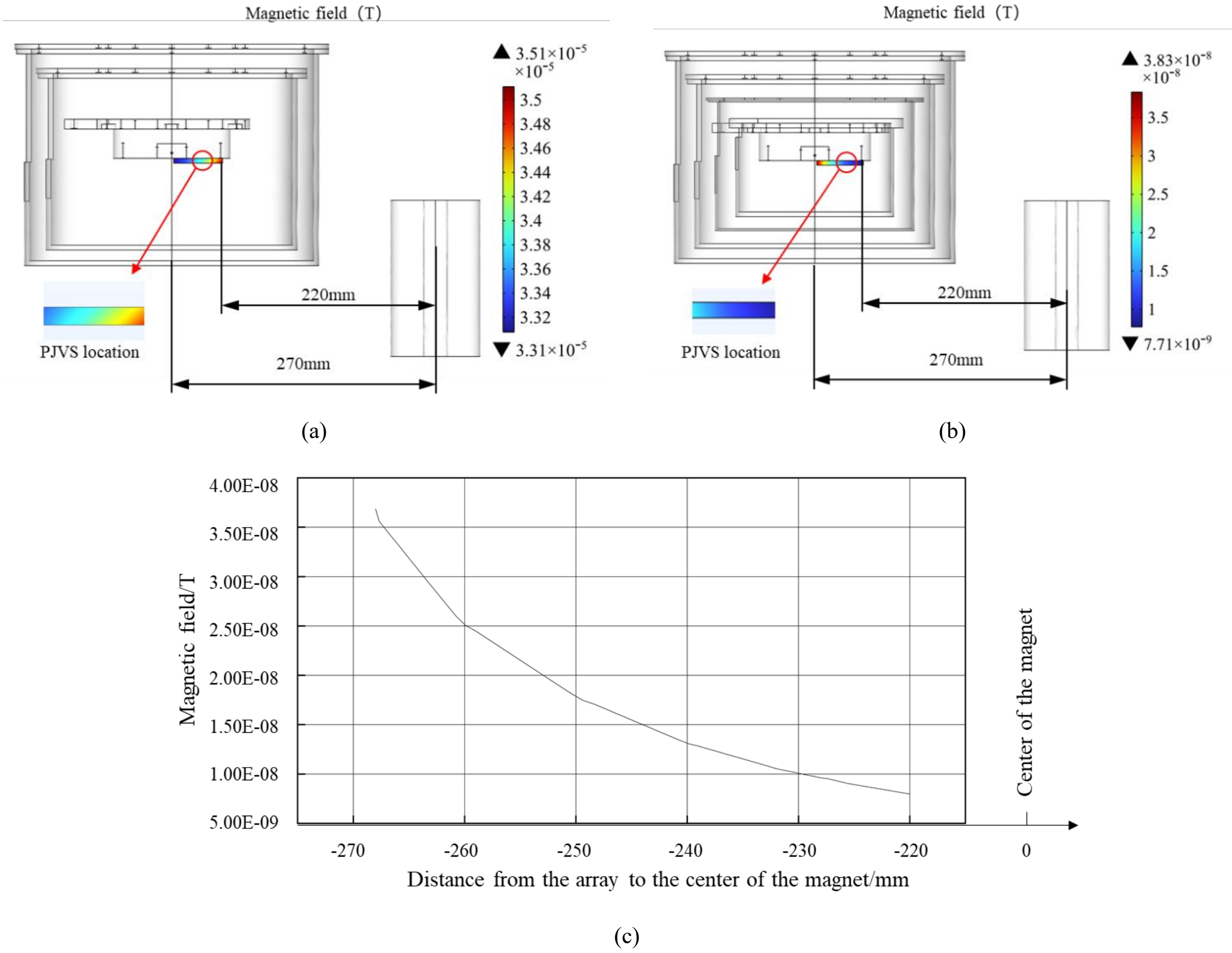

**Extended Data Fig. 1 | Finite-element characterization of multilayer magnetic shielding performance for the PJVS operating region.** For conservative simulation evaluation, the PJVS functional region is approximated as a rectangular prism (red region in Fig. M2) with dimensions slightly larger than the actual device footprint, enabling worst-case stray field assessment. **(a)** Simulated magnetic field distribution with only two outer ferromagnetic shielding layers activated. The stray field originating from the 6 T superconducting magnet is attenuated from approximately 25 mT to below $3.5\times 10^{-5}$ T within the PJVS region. **(b)** Magnetic field distribution under full multilayer shielding integration, comprising ferromagnetic, permalloy, and superconducting niobium layers. The residual field is further suppressed below $3.8 \times 10^{-8}$ T (38 nT), achieving an additional attenuation of over three orders of magnitude. **(c)** Horizontal magnetic field profile across the top surface of the PJVS operating region, with the origin defined at the central axis of the superconducting magnet. The results confirm that the magnetic field across the entire PJVS functional area remains below 50 nT under full magnet excitation, validating the reliability and efficacy of the multilayer cascaded shielding architecture.

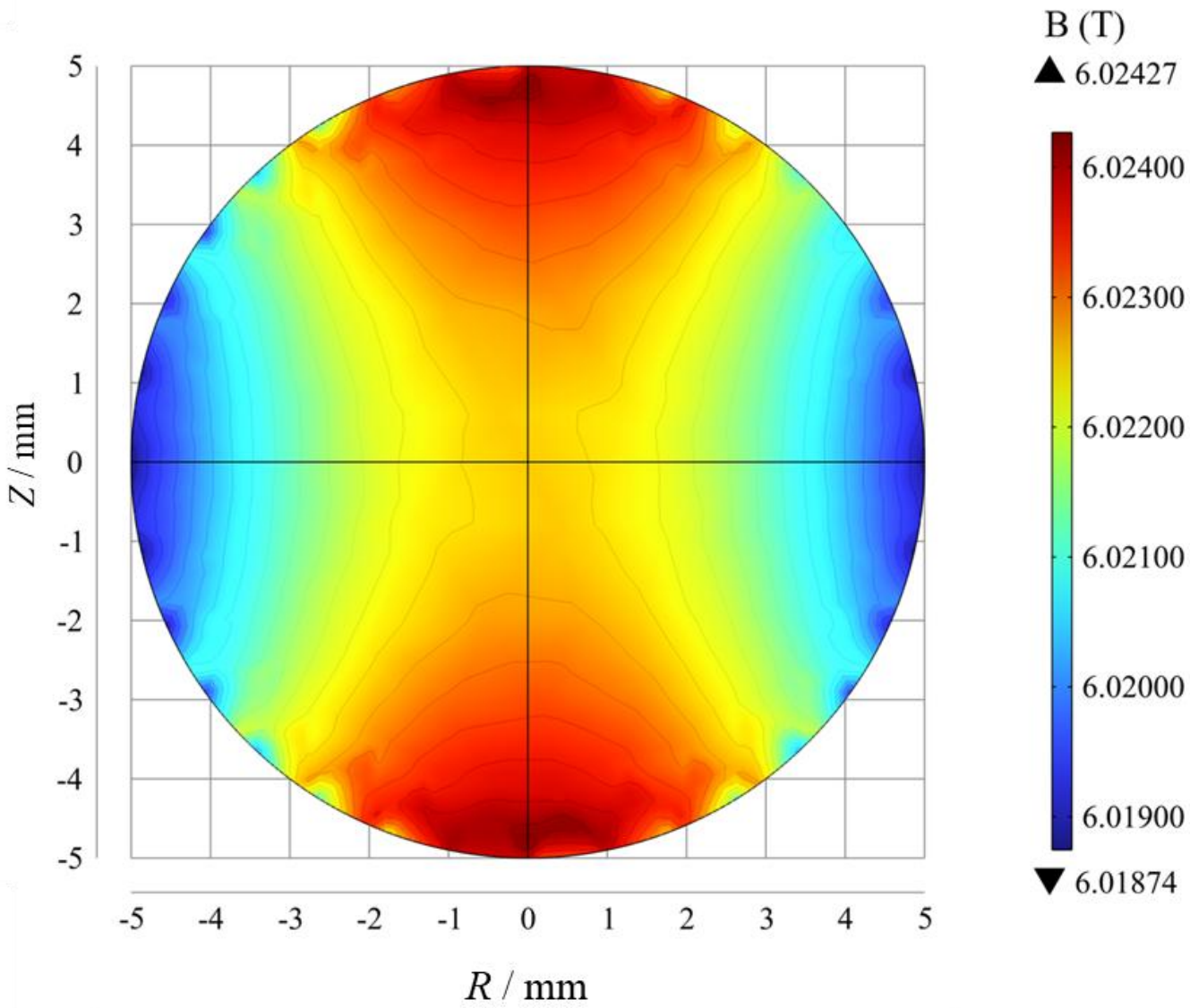

**Extended Data Fig. 2 | Magnetic field distribution and spatial uniformity of the superconducting magnet within the 10 mm diameter spherical volume (DSV).** Numerical simulation confirms a field uniformity better than ±0.05% across the 10 mm DSV designated for QHR device placement. The maximum magnetic load on the inner and outer magnet coils reaches 6.01 T and 5.15 T, corresponding to operational safety margins of 30% and 29.4%, respectively. These results verify that the customized magnet design delivers sufficient field uniformity and stable operating margins to support reliable quantum Hall quantization under integrated system operating conditions.

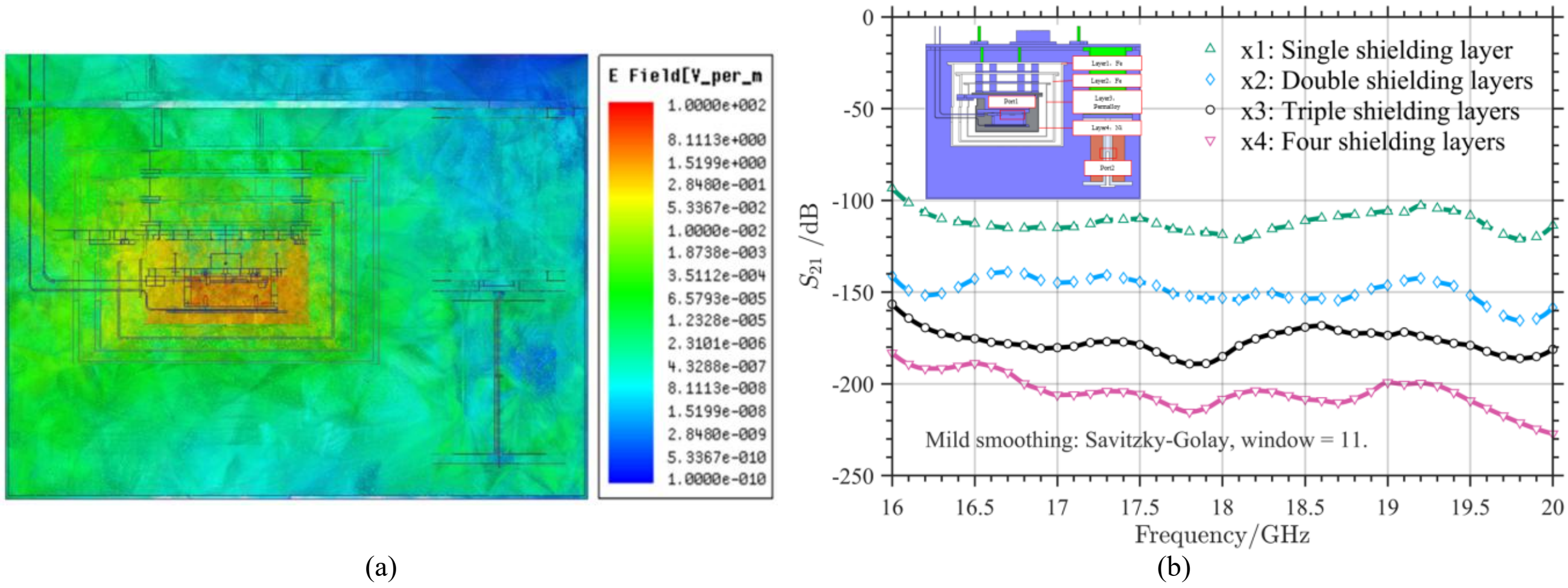

**Extended Data Fig. 3 | Microwave shielding performance of the multilayer assembly.**

Full-wave electromagnetic simulations are performed at a PJVS operating power of 30 dBm and a central frequency of 16~20 GHz to evaluate microwave crosstalk toward the QHR subsystem. **(a)** Three-dimensional electric field distribution simulation reveals substantial microwave attenuation within the shielding cavity, with negligible field intensity detected in the QHR functional region. **(b)** Evolution of $S_{21}$ transmission with progressive activation of the shielding layers from the single shielded case to the full multilayer configuration, demonstrating monotonic improvement in shielding effectiveness. These results quantitatively demonstrate that the continuous multilayer shielding structure provides cumulative attenuation of microwave propagation, enabling effective suppression of PJVS-induced electromagnetic interference to the QHR.

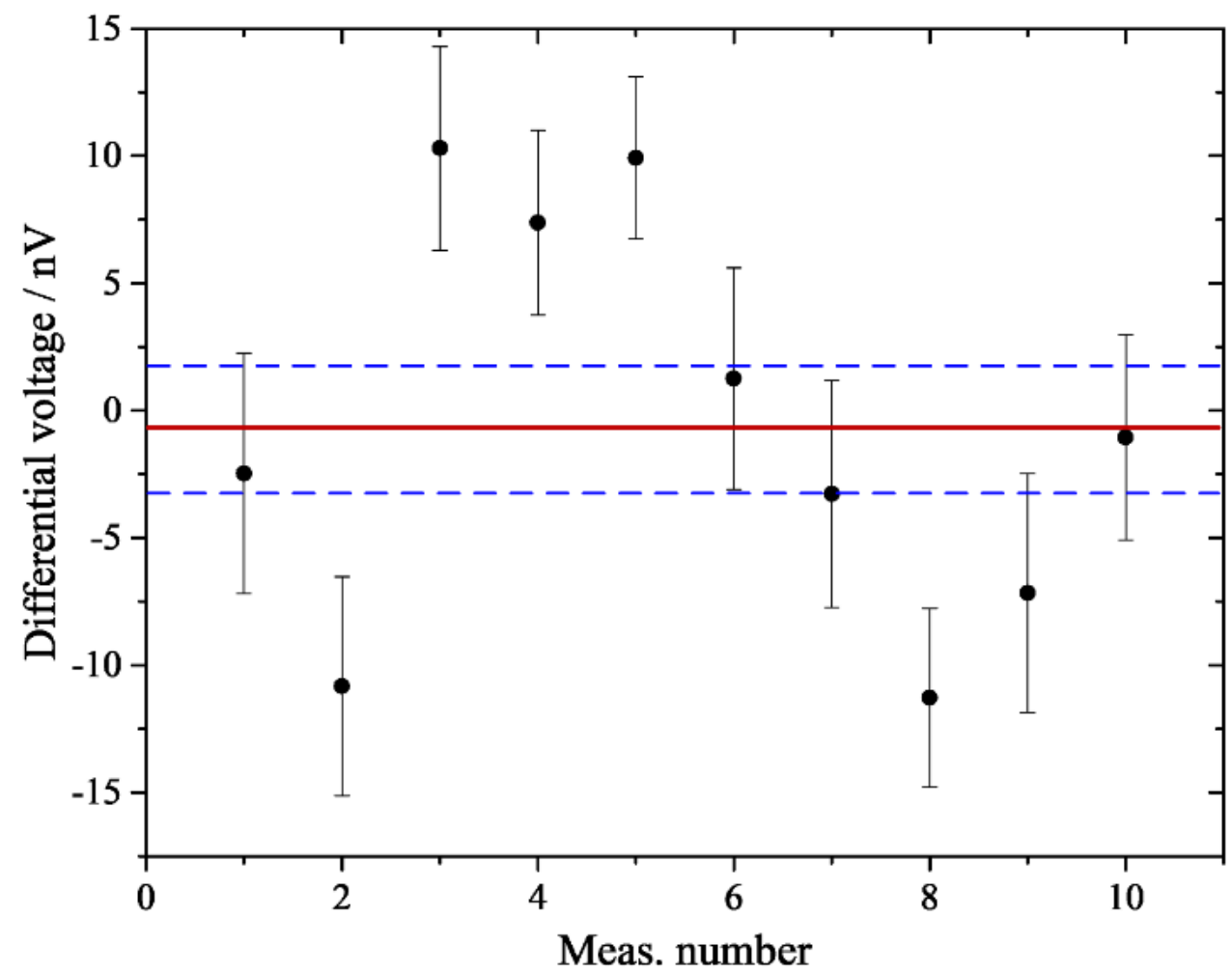

**Extended Data Fig. 4 | Statistical analysis of voltage consistency between the integrated PJVS and an independent reference PJVS system.** Voltage differences are acquired from ten consecutive comparison measurements at ±2 V output states, following the differential measurement protocol detailed in the Methods. Each data point represents the voltage deviation between the integrated and reference PJVS systems. The mean deviation is -0.7 nV (red solid line), with a statistical standard deviation of 2.5 nV (blue dashed lines), yielding relative measurement dispersion at the $10^{-9}$ level. These statistical results demonstrate that the integrated PJVS maintains precision and stability comparable to conventional standalone quantum voltage standards, with no measurable performance degradation induced by system-level co-integration.

## References

[1] Bureau International des Poids et Mesures, The International System of Units 9th edn (2019), https://www.bipm.org/en/publications/si-brochure.

[2] Michael Stock, et al. The revision of the SI—the result of three decades of progress in metrology. Metrologia 56, 022001 (2019).

[3] B D Josephson. Possible new effects in superconductive tunnelling, Phys. Lett. 1, 251 (1962).

[4] K von Klitzing, et al. New method for high-accuracy determination of the fine-structure constant based on quantized Hall resistance. Phys. Rev. Lett. 45, 494-497 (1980).

[5] H Pothier, et al. Single-electron pump based on charging effects. Europhysics Letters 17, 249-253 (1992).

[6] Scherer H and Camarota B. Quantum metrology triangle experiments: a status review. Meas. Sci. Technol. 23, 124010 (2012).

[7] Keller M W. Current status of the quantum metrology triangle. Metrologia 45, 102–109 (2008).

[8] Taylor B and Witt T. New international electrical reference standards based on the Josephson and quantum Hall effects. Metrologia 26, 47–62 (1989).

[9] Lafont F, et al. A programmable quantum current standard from the Josephson and the quantum Hall effects. Journal of Applied Physics 115, 044509 (2014).

[10] Djordjevic S, et al. Improvements of the programmable quantum current generator for better traceability of electrical current measurements. Metrologia 58, 045005 (2021).

[11] Chae D H, et al. Series connection of quantum Hall resistance array and programmable Josephson voltage standard for current generation at one microampere. Metrologia 59, 065011 (2022).

[12] Kaneko N H, et al. Perspectives of the generation and measurement of small electric currents. Meas. Sci. Technol. 35, 011001 (2024).

[13] C A Hamilton, et al. Josephson D/A converter with fundamental accuracy. IEEE Trans. Instrum. Meas. 44, 223–225 (1995).

[14] A Rüfenacht, et al. Dual-frequency-bias programmable Josephson voltage standard circuit. IEEE Trans. Instrum. Meas. 74, 2006007 (2025).

[15] Dresselhaus P D, et al. 10 V programmable Josephson voltage standard circuits using NbSi-barrier junctions. IEEE Trans. Appl. Supercond. 21, 693–6 (2011).

[16] Mueller F, et al. 1 V and 10 V SNS programmable voltage standards for 70 GHz. IEEE Trans. Appl. Supercond. 19, 981–6 (2009).

[17] Mueller F, et al. NbSi barrier junctions tuned for metrological applications up to 70 GHz: 20 V arrays for programmable Josephson voltage standards. IEEE Trans. Appl. Supercond. 23, 1101005 (2013).

[18] Yamamori H, et al. A 10 V programmable Josephson voltage standard circuit with a maximum output voltage of 20 V. Supercond. Sci. Technol. 21, 105007 (2008).

[19] Bruno Trinchera, et al. Development of a PJVS system for quantum-based sampled power measurements. Measurement 219, 113275 (2023).

[20] Benz S P and Hamilton C A. A pulse-driven programmable Josephson voltage standard. Appl. Phys. Lett. 68, 3171 (1996).

[21] Benz S P, et al. 1 V Josephson arbitrary waveform synthesizer. IEEE Trans. Appl. Supercond. 25, 1–8 (2015).

[22] Benz S P, et al. Performance improvements for the NIST 1 V Josephson arbitrary waveform synthesizer IEEE Trans. Appl. Supercond. 25, 1–5 (2015).

[23] N E Flowers-Jacobs, et al. Development and applications of a four-volt Josephson arbitrary waveform synthesizer. ISEC, Riverside, CA, USA, 1-2 (2019).

[24] Zhou K, et al. Four-volt Josephson arbitrary waveform synthesizer at NIM. CPEM, Madrid, Spain, 1-2 (2026).

[25] Flowers-Jacobs N E, et al. Two-volt Josephson arbitrary waveform synthesizer using wilkinson dividers. IEEE Trans. Appl. Supercond. 26, 1–7 (2016).

[26] Zhou K, et al. Zero-compensation method and reduced inductive voltage error for the AC Josephson voltage standard IEEE Trans. Appl. Supercond. 25, 1400806 (2015).

[27] Kieler O F, et al. Towards a 1 V Josephson arbitrary waveform synthesizer. IEEE Trans. Appl. Supercond. 25, 1–5 (2015)

[28]Behr R, et al. Direct comparison of a 1 V Josephson arbitrary waveform synthesizer and an ac quantum voltmeter. Metrologia 52, 528–37 (2015).

[29]Rüfenacht Alain, et al. Impact of the latest generation of Josephson voltage standards in ac and dc electric metrology. Metrologia 55, S152-S173 (2018).

[30]Bauer S, et al. Josephson voltage standards as toolkit for precision metrological applications at PTB. Meas. Sci. Technol. 34, 032001 (2023).

[31]L. Howe, et al. Cryogen-free operation of 10 V programmable Josephson voltage standards. IEEE Trans. Appl. Supercond. 23, 1300605 (2013).

[32]van den Brom H E, et al. AC–DC calibrations with a pulse-driven AC Josephson voltage standard operated in a small cryostat. IEEE Trans. Instrum. Meas. 66, 1391–6 (2017).

[33]Clothier W, et al. A determination of the volt. Metrologia 26, 9–46 (1989).

[34]Benz S P, et al. AC and DC bipolar voltage source using quantized pulses. IEEE Trans. Instrum. Meas. 48, 266–9 (1999).

[35]Filipski P S, et al. International comparison of quantum AC voltage standards for frequencies up to 100 kHz. Measurement 45, 2218–25 (2012).

[36]Jinni Lee, et al. An ac quantum voltmeter based on a 10 V programmable Josephson array. Metrologia 50, 612 (2013).

[37]Schubert M, et al. An AC Josephson voltage standard up to the kilohertz range tested in a calibration laboratory. IEEE Trans. Instrum. Meas. 64, 1620–6 (2015).

[38]Ralf Behr and Luis Palafox. An AC quantum voltmeter for frequencies up to 100 kHz using sub-sampling. Metrologia 58, 025010 (2021).

[39]Kurten Ihlenfeld W G and Pinheiro Landim R. An automated Josephson-based AC-voltage calibration system. IEEE Trans. Instrum. Meas. 64, 1779–84 (2015).

[40]Lipe T E, et al. Thermal voltage converter calibrations using a quantum ac standard. Metrologia 45, 275–80 (2008).

[41]Jason M Underwood. Uncertainty analysis for ac–dc difference measurements with the AC Josephson voltage standard. Metrologia 56, 015012 (2019).

[42]E Luna et al. Seamless sampling method to characterize multitone signals using the AC quantum voltmeter. Meas. Sci. Technol. 36, 106133 (2025).

[43]D Georgakopoulos, et al. Josephson arbitrary waveform synthesizer as a reference standard for the measurement of the phase of harmonics in distorted waveforms. IEEE Trans. Instrum. Meas. 68, 1927-1934 (2019).

[44]Maruyama M, et al. Calibration system for zener voltage standards using a 10 V programmable Josephson voltage standard at NMIJ. IEEE Trans. Instrum. Meas. 64, 1606–12 (2015).

[45]Maruyama M, et al. Evaluation of linearity characteristics in digital voltmeters using a PJVS system with a 10 K cooler. IEEE Trans. Instrum. Meas. 64, 1613–9 (2015).

[46]Overney F, et al. Characterization of metrological grade analog-to-digital converters using a programmable Josephson voltage standard. IEEE Trans. Instrum. Meas. 60, 2172-2177 (2011).

[47]B C Waltrip, et al. AC power standard using a programmable Josephson voltage standard. IEEE Trans. Instrum. Meas. 58, 1041–1048 (2009).

[48]B V Djokic. Low-frequency quantum-based AC power standard at NRC Canada. IEEE Trans. Instrum. Meas. 62, 1699-1703 (2013).

[49]L Palafox, et al. The Josephson-effect-based primary AC power standard at the PTB: Progress report. IEEE Trans. Instrum. Meas. 58, 1049-1053 (2009).

[50]L Qian, et al. Development of DC power standard source based on programmable Josephson voltage standard. IEEE Trans. Instrum. Meas. 74, 1505611 (2025).

[51]Bruno Trinchera, et al. Quantum sampling AC standard for electrical power metrology based on programmable Josephson junction series array. Measurement 233, 114747 (2024).

[52]B Waltrip, et al. A sampling wattmeter with extended frequency range. IEEE Trans. Instrum. Meas. 68, 2187-2194 (2019).

[53]B C Waltrip, et al. Comparison of AC power referenced to either PJVS or JAWS. IEEE Trans. Instrum. Meas. 70, 1-6 (2021).
[54]B C Waltrip, et al. A comparison between the NIST PJVS-based power standard and the NRC current-comparator-based power standard. IEEE Trans. Instrum. Meas. 64, 14-18 (2015).
[55]Frédéric Overney and Blaise Jeanneret. Impedance bridges: from Wheatstone to Josephson. Metrologia 55, S119 (2018).
[56]Frédéric Overney, et al. Dual Josephson impedance bridge: towards a universal bridge for impedance metrology. Metrologia 57, 065014 (2020).
[57]Lee J, et al. The Josephson two-terminal-pair impedance bridge. Metrologia 47, 453–9 (2010).
[58]Lee J, et al. Programmable Josephson arrays for impedance measurements. IEEE Trans. Instrum. Meas. 60, 2596–601 (2011).
[59]Hagen T, et al. A Josephson impedance bridge based on programmable Josephson voltage standards. IEEE Trans. Instrum. Meas. 66, 1539–45 (2017).
[60]Bauer S, et al. A novel two-terminal-pair pulse-driven Josephson impedance bridge linking a 10 nF capacitance standard to the quantized Hall resistance. Metrologia 54, 152–60 (2017).
[61]Overney F, et al. Josephson-based full digital bridge for high-accuracy impedance comparisons. Metrologia 53, 1045–53 (2016).
[62]Bauer S et al. A four-terminal-pair Josephson impedancebridge combined with a graphene-quantized Hall resistance. Meas. Sci. Technol. 32, 065007 (2021).
[63]K V Klitzing and G. Ebert. Application of the quantum Hall effect in metrology. Metrologia 21, 11–18 (1985).
[64]Jeckelmann B and Jeanneret B. The quantum Hall effect as an electrical resistance standard. Rep. Prog. Phys. 64, 1603–1655 (2001).
[65]Hartland A, et al. Direct comparison of the quantized Hall resistance in gallium arsenide and silicon. Phys. Rev. Letters 66, 969-973 (1991).
[66]Jeckelmann B, et al. High-precision measurements of the quantized Hall resistance: Experimental conditions for universality. Phys. Rev. B. 55, 13124-13134 (1997).
[67]Poirier W and Schopfer F. Resistance metrology based on the quantum Hall effect. Eur. Phys. J. Spec. Top. 172, 207–245 (2009).
[68]Schopfer F and Poirier W. Quantum resistance standard accuracy close to the zero-dissipation state. J. Appl. Phys. 114, 064508 (2013).
[69]Zhang Y, et al. Experimental observation of the quantum Hall effect and Berry's phase in graphene. Nature 438, 201–204 (2005).
[70]Giesbers A J M, et al. Quantum resistance metrology in graphene. Appl. Phys. Lett. 93, 222109 (2008).
[71]Satrapinski A, et al. Precision quantum Hall resistance measurement on epitaxial graphene device in low magnetic field. Appl. Phys. Lett. 103, 173509 (2013).
[72]A F Rigosi and R E Elmquist. The quantum Hall effect in the era of the new Si. Semicond. Sci. Technol. 34, 093004 (2019).
[73]A Tzalenchuk. Towards a quantum resistance standard based on epitaxial graphene. Nature Nanotechnol. 5, 186–189 (2010).
[74]R Ribeiro-Palau. Quantum Hall resistance standard in graphene devices under relaxed experimental conditions. Nature Nanotechnol. 10, 965–971 (2015).
[75]M Kruskopf and R E Elmquist. Epitaxial graphene for quantum resistance metrology. Metrologia 55, R27–R36 (2018).
[76]S Novikov, et al. Fabrication and study of large-area QHE devices based on epitaxial graphene. IEEE Trans. Instrum. Meas. 64, 1533–1538 (2015).
[77]M Kruskopf. Graphene quantum Hall effect devices for AC and DC electrical metrology. IEEE Trans. Electron Devices 68, 3672–3677 (2021).
[78]T J B M Janssen. Precision comparison of the quantum Hall effect in graphene and gallium arsenide. Metrologia 49, 294–306 (2012).
[79]T Oe. Comparison between NIST graphene and AIST GaAs quantized Hall devices. IEEE Trans. Instrum. Meas. 69, 3103–3108 (2020).
[80]H He. Accurate graphene quantum Hall arrays for the new international system of units. Nature Commun. 13, 1–9 (2022).
[81]A R Panna. Graphene quantum Hall effect parallel resistance arrays. Phys. Rev. B, Condens. Matter 103, 075408 (2021).

[82]Chatterjee A, et al. Performance and stability assessment of graphene-based quantum Hall devices for resistance metrology. IEEE Trans. Instrum. Meas. 72, 1502206 (2023).

[83]Chae D H, et al. Recent advances and perspectives in quantum electrical metrology with epitaxial graphene. Meas. Sci. Technol. 37, 151001 (2026).

[84]Sergiy Rozhko, et al. Carrier concentration adjustment in graphene QHR devices for metrology applications. Journal of Applied Physics 138, 134503 (2025).

[85]Yefei Yin, et al. Graphene quantum Hall resistance standard for realizing the unit of electrical resistance under relaxed experimental conditions. Physical Review Applied 23, 014025 (2025).

[86]Yefei Yin, et al. Quantum Hall resistance standards based on epitaxial graphene with p-type conductivity. Applied Physics Letters, 125, 064001 (2024).

[87]Giblin, S. P. et al. Towards a quantum representation of the ampere using single electron pumps. Nat. Commun. 3, 930 (2012).

[88]Fujiwara A, et al. Silicon quantum dot single-electron pumps for the closure of the quantum metrology triangle. Electrochemical Society Transactions, 112, 119-130 (2023).

[89]Scherer H and Schumacher H W. Single-electron pumps and quantum current metrology in the revised SI. Annalen der Physik 531, 1800371 (2019).

[90]Kaneko N H, et al. A review of the quantum current standard. Meas. Sci. Technol. 27, 032001 (2016).

[91]Stein F, et al. Validation of a quantized-current source with 0.2 ppm uncertainty. Appl. Phys. Lett. 107, 103501 (2015).

[92]Stein F, et al. Robustness of single-electron pumps at sub-ppm current accuracy level. Metrologia 54, S1 (2017).

[93]J P Pekola, et al. Single-electron current sources: Toward a refined definition of the ampere. Rev. Mod. Phys. 85, 1421 (2013).

[94]B Kaestner and V Kashcheyevs. Non-adiabatic quantized charge pumping with tunable-barrier quantum dots: A review of current progress, Rep. Prog. Phys. 78, 103901 (2015).

[95]Brun-Picard J, et al. Practical quantum realization of the ampere from the elementary charge. Physical Review X 6, 041051 (2016).

[96]S. Nakamura, et al. Single-electron pumping by parallel SINIS turnstiles for quantum current standard. IEEE Trans. Instrum. Meas. 64, 1696-1701 (2015).

[97]Djordjevic S, et al. A primary quantum current standard based on the Josephson and the quantum Hall effects. Nature Communications, 16, 1447 (2025).

[98]Poirier W and Djordjevic S. A universal quantum electrical standard is getting closer. Nature Electronics 8, 632-634 (2025).

[99]Hohls F, et al. Semiconductor quantized voltage source. Physical Review Letters, 109, 056802 (2012).

[100] Rodenbach L K, et al. A unified realization of electrical quantities from the quantum International System of Units. Nature Electronics, 8, 663-671 (2025).

[101] Patel D K, et al. A zero external magnetic field quantum standard of resistance at the $10^{-9}$ level. Nature Electronics, 7, 1111-1116 (2024).

[102] A voltage-balanced device for quantum resistance metrology. Nat Electron 7, 436–437 (2024).

[103] Nathaniel J Huáng, et al. Quantum anomalous Hall effect for metrology. Applied Physics Letters 126, 040501 (2025).

[104] Chen S, et al. Development of an 8-T conduction-cooled superconducting magnet with 300-mm warm bore for material processing application. IEEE Trans. Appl. Supercond. 24, 4701605 (2014).

[105] Zhang Z, et al. Engineering-based design and fabrication procedure for mid-temperature REBCO magnets accommodating the strong Ic anisotropy. Superconductivity 1, 100005 (2022).

[106] Ni Z, et al. A homogeneous superconducting magnet design using a hybrid optimization algorithm. Meas. Sci. Technol. 24, 125402 (2013).

[107] Li Y, et al. Shape optimization of ferromagnetic pole of a ferromagnetic-superconducting MRI magnet. IEEE Trans. Appl. Supercond. 26, 4404005 (2016).

[108] Filippidis S P, et al. Overview of the electromagnetic optimization literature of superconducting solenoidal magnets and coils. IEEE Trans. Appl. Supercond. 33, 4901221 (2023).

[109] Tomassetti G, et al. Direct and surrogate optimization in applied superconductivity: state of the art, perspectives and challenges. Supercond. Sci. Technol. 38, 073001 (2025).

[110] Wang Y, et al. A novel passive shimming scheme using explicit control of magnetic field qualities with minimal use of ferromagnetic materials. Magn. Reson. Med. 88, 2732–2744 (2022).

[111] Wang Y, et al. Accurate magnetization modeling in multi-dimensional applications. J. Appl. Phys. 135, 023901 (2024).

[112] Wang L, et al. Design of the permanent magnet diverter for deflecting electrons on wide-field x-ray telescope. IEEE Trans. Appl. Supercond. 30, 3602405 (2020).

[113] Niu C, et al. Numerical analysis of eddy current induced by z-gradient coil in a superconducting MRI magnet. IEEE Trans. Appl. Supercond. 30, 4400606 (2020).

[114] Qu H, et al. Design of the superconducting main magnet based on variable density method in cylindrical MRI scanner. IEEE Trans. Appl. Supercond. 30, 4400805 (2020).

[115] Wang L, et al. A novel method to eliminate the screening current–induced magnetic field in a non-insulated REBCO double pancake coil. J Supercond Nov Magn 33, 1729–1735 (2020).

[116] Niu F, et al. Effect of magnetic shielding on levitation characteristics of superconducting gravimeter. IEEE Trans. Appl. Supercond. 32, 3800414 (2022).

[117] Hinterberger A, et al. Superconducting shielding with Pb and Nb tubes for momentum sensitive measurements of neutral antimatter. J. Instrum. 12, T09002 (2017).

[118] Dickerson S, et al. A high-performance magnetic shield with large length-to-diameter ratio. Rev. Sci. Instrum. 83, 065108 (2012).

[119] Bork J, et al. The 8-layered magnetically shielded room of the PTB: Design and construction. Proc. 12th Int. Conf. Biomagnetism, 970 − 973 (2001).

[120] Altarev I, et al. A magnetically shielded room with ultra low residual field and gradient. Rev. Sci. Instrum. 85, 075106 (2014).

[121] C J Burroughs, et al. NIST 10 V programmable Josephson voltage standard system. IEEE Trans. Instrum. Meas. 60, 2482–2488 (2011).